\documentclass[final,3p,times]{elsarticle}

\usepackage{amssymb}
\usepackage{graphicx}
\usepackage{amsfonts}
\usepackage{amsmath}

\begin{document}

\begin{frontmatter}



\title{Study of invariance of nonextensive statistics under the uniform energy spectrum translation}


\author{A.S.~Parvan}

\affiliation{organization={Bogoliubov Laboratory of Theoretical Physics, Joint Institute for Nuclear Research},
            city={Dubna},
            postcode={141980},
            country={Russia}}
\affiliation{organization={Department of Theoretical Physics, Horia Hulubei National Institute of Physics and Nuclear Engineering},
            city={Bucharest-Magurele},
            country={Romania}}

\begin{abstract}
The general formalisms of the $q$-dual statistics, the Boltzmann-Gibbs statistics, and three versions of the Tsallis statistics known as Tsallis-1, Tsallis-2, and Tsallis-3 statistics have been considered in the canonical ensemble. We have rigorously proved that the probability distribution of the Tsallis-1 statistics is invariant under the uniform energy spectrum translation at a fixed temperature. This invariance demonstrates that the formalism of the Tsallis-1 statistics is consistent with the fundamentals of the equilibrium statistical mechanics. The same results we have obtained for the probability distributions of the Tsallis-3 statistics, Boltzmann-Gibbs statistics, and $q$-dual statistics. However, we have found that the probability distribution of the Tsallis-2 statistics, the expectation values of which are not consistent with the normalization condition of probabilities, is indeed not invariant under the overall shift in energy as expected.
\end{abstract}




\end{frontmatter}


\section{Introduction}
In modern physics, the microscopic foundation of the statistical mechanics' paradigm based on the nonextensive entropy measure~\cite{Tsal88} is a primordial task. It is well-known that one of the fundamental aspects of the theory of the usual Boltzmann-Gibbs statistical mechanics is the invariance of the probability distribution under the uniform translation of energy spectrum. This invariance is a consequence of the first law of equilibrium thermodynamics, which in the case of a cyclic process, may be written in the form of the integral of energy change~\cite{Prigogine}:
\begin{equation}\label{i}
  \oint dU = 0.
\end{equation}
In thermodynamics, the internal energy $U$ is a state function and the infinitesimal change $dU$ of the internal energy associated with any infinitesimal transformation of state is a total differential~\cite{Greiner}. The total change in energy $U$ is independent of the transformation path. It depends only on the initial and final states. Thus, in thermodynamics, the internal energy $U$ can be only defined up to an arbitrary additive constant $U_{0}$~\cite{Prigogine}:
\begin{equation}\label{ii}
  U= U(T,V,N) + U_{0}.
\end{equation}
Then, the invariance of the probability distribution under the uniform energy spectrum translation may be considered as a statement of the first law of thermodynamics~(\ref{i}) and additivity~(\ref{ii}).

Nowadays, there are at least three versions of the Tsallis statistics~\cite{Tsal88,Tsal98} based on the same generalized entropy~\cite{Tsal88,Havrda,Daroczy,Wehrl} (see more concrete explanations in ref.~\cite{Tsal_book}) that differ from each other only in the definition of the expectation values of operators. The first variant of the Tsallis statistics~\cite{Tsal88,Tsal98,Parvan2006a}, also called the Tsallis-1 statistics, is defined by the standard expectation values of operators as in the Boltzmann-Gibbs statistics. Such expectation values are consistent with the normalization condition of probabilities in total agreement with the statistical mechanics and theory of probability requirements. The second variant of the Tsallis statistics (the Tsallis-2 statistics)~\cite{Tsal88,Tsal98,Curado91} is based on the generalized expectation values of operators, which are not consistent with the normalization condition of probabilities. Such unconventional expectation values lead to an inconsistent connection between the statistical mechanics, probability theory, and theory of equilibrium thermodynamics. In the third variant of the Tsallis statistics~\cite{Tsal98}, also called the Tsallis-3 statistics, the normalized generalized expectation values of operators are used. However, in contrast with the Tsallis-2 statistics, the normalized generalized expectation values of the Tsallis-3 statistics are consistent with the normalization condition of probabilities of microstates.

The present paper is motivated by the uncertainties dominant in high-energy physics concerning the correct variant of the Tsallis statistics. Generally, in high-energy physics, the phenomenological Tsallis distribution~\cite{Cleymans2012,Cleymans12a,Bediaga00,Beck00} and different Tsallis-like distributions~\cite{Cleymans09,Wong15,Shen18} of the power-law form inspired by the Tsallis statistics~\cite{Tsal88} are used to describe the experimental data for the transverse momentum distribution of particles produced in proton-proton and heavy-ion collisions at the LHC and RHIC energies~\cite{STAR,PHENIX1,ALICE_charged,ALICE_piplus,ALICE_PbPb,CMS1,ATLAS,ALICE_deuteron,CMS2}. These distributions are the single-particle distribution functions, which have difficulties of the fundamental character in the Tsallis statistics (see the detailed proof in refs.~\cite{Parvan2017a,Parvan2020a,Parvan2021a}). Nevertheless, such difficulties for the phenomenological Tsallis distribution are eliminable in the framework of the $q$-dual statistics introduced in ref.~\cite{Parvan2020b}. In high-energy physics, the consistent single-particle distribution functions were found only for the Tsallis-1 statistics~\cite{Parvan2017a,Parvan2020a}. However, in the high-energy physics community, only the Tsallis-3 statistics is most frequently considered correct since it is hypothesized that only the probability distribution of the Tsallis-3 statistics is invariant under the uniform energy spectrum translation (see, for example, the reference~\cite{Kapusta}). In this paper, we are going to show that the Tsallis-1 statistics is also invariant under the uniform energy spectrum translation.

The conviction that only the Tsallis-3 statistics is invariant under the uniform energy spectrum translation has its origins in Tsallis' works. In ref.~\cite{Tsal_book}, it was stated that the probability distribution of the Tsallis-1 statistics is not invariant under the overall shift in energy at fixed Lagrange multiplier $\beta$ associated with the energy, i.e., the thermodynamical quantities as functions of $\beta$ depend on the choice of the origin of energies. In refs.~\cite{Tsal98,Tsal_book}, the same statement was given for the Tsallis-2 statistics. However, in the same refs.~\cite{Tsal98,Tsal_book}, it was shown that the probability distribution of the Tsallis-3 statistics is invariant under the overall shift in energy at fixed Lagrange multiplier $\beta$. Based on these conclusions, in refs.~\cite{Tsal_book,Kapusta}, it is stated that of all three formulations of the Tsallis statistics only the Tsallis-3 statistics is consistent with the fundamentals of the equilibrium statistical mechanics. However, this statement contradicts the results of the earlier work~\cite{Sisto} in which it was demonstrated that the probability distribution of the Tsallis-1 statistics is also invariant under the uniform energy spectrum translation.

In the present paper, in comparison with ref.~\cite{Sisto}, we introduce a more general and explicit method to demonstrate the invariance of the probability distributions under the uniform energy spectrum translation. Using this new method, we strongly prove that the probability distribution of the Tsallis-1 statistics is invariant under the overall shift in energy at a fixed temperature. This important property is found in the framework of the complete formalism of the Tsallis-1 statistics obtained in ref.~\cite{Parvan2006a} and in present paper. We also prove the invariance of the probability distributions of the Tsallis-3 statistics, the $q$-dual statistics, and Boltzmann-Gibbs statistics. However, we demonstrate that the probability distribution of the Tsallis-2 statistics is not invariant under the energy shift, even in terms of temperature.

The paper is organized as follows. In Secs.~\ref{sec1} -- \ref{sec5} we discuss the general formalisms of the Boltzmann-Gibbs statistics, Tsallis-1 statistics, Tsallis-2 statistics, Tsallis-3 statistics, and $q$-dual statistics, respectively, and study their invariance under the energy shift. The discussions and summary are given in Sec.~\ref{sec6}.

\section{Boltzmann-Gibbs statistics}\label{sec1}
The Boltzmann-Gibbs statistics is defined by the Boltzmann-Gibbs entropy with the probabilities $p_{i}$ of the microstates of the system normalized to unity
\begin{align}\label{1}
 S = &  -\sum_{i} p_{i} \ln p_{i},  \\ \label{2}
 1 = & \sum\limits_{i} p_{i}
\end{align}
and by the standard expectation values of an operator $A$
\begin{equation}\label{3}
   \langle A \rangle = \sum\limits_{i} p_{i} A_{i},
\end{equation}
where $A_{i}=\langle i | A | i \rangle$ is the value of operator $A$ in the $i$-th microscopic state of the system. Here and throughout the paper we use the system of natural units $\hbar=c=k_{B}=1$.

Let us consider the canonical ensemble. The thermodynamic potential of the canonical ensemble, the free energy $F$, is the Legendre transform of the fundamental thermodynamic potential, the energy $E$:
\begin{align}\label{4}
 F = & E -TS =  \sum\limits_{i}  p_{i} (E_{i} + T \ln p_{i}), \\ \label{5}
 E = & \sum_{i}  p_{i} E_{i},
\end{align}
where $E$ is the mean energy and $E_{i}$ is the energy of the $i$-th microscopic state of the system.

The unknown probabilities $\{p_{i}\}$ are found from the first and second laws of thermodynamics (the principle of maximum entropy). This means that in the canonical ensemble the set of equilibrium probabilities $\{p_{i}\}$ can be derived from the constrained local extremum of the thermodynamic potential (\ref{4}) by the method of the Lagrange multipliers (see, for example, Refs.~\cite{Jaynes2,Krasnov,Parvan2015}):
\begin{align}\label{6}
 \Phi = & F - \lambda \varphi,  \\ \label{7}
    \varphi= & \sum\limits_{i} p_{i} - 1 = 0, \\ \label{8}
  \frac{\partial \Phi}{\partial p_{i}} = & 0,
\end{align}
where $\lambda$ is an arbitrary real constant.

Substituting Eq.~(\ref{4}) into Eqs.~(\ref{6})--(\ref{8}), we obtain
\begin{equation}\label{9}
p_{i} = e^{\frac{\Lambda-E_{i}}{T}}.
\end{equation}
where $\Lambda\equiv \lambda-T$ and $\partial E_{i}/\partial p_{i}=0$. Substituting Eq.~(\ref{9}) into Eq.~(\ref{4}) and using Eq.~(\ref{2}), we obtain
\begin{equation}\label{10}
  F=\Lambda =\lambda-T
\end{equation}
and~\cite{Zubarev}
\begin{align}\label{11}
  p_{i} = & e^{\frac{F-E_{i}}{T}} = e^{-S+\frac{E-E_{i}}{T}} \equiv \frac{1}{Z} e^{-\frac{E_{i}}{T}}, \\ \label{12}
  Z = &  \sum\limits_{i} e^{-\frac{E_{i}}{T}}, \\ \label{13}
  F = & -T \ln Z,
\end{align}
where $Z$ is the partition function.

Let us prove that the probability distribution (\ref{11}) is invariant under an overall shift in energy of microstates
\begin{equation}\label{14}
  E'_{i}=E_{i} + E_{0},
\end{equation}
where $E_{0}$ is the same constant for all microstates of the system. Let us suppose that the probability (\ref{11}) is invariant under the transformation (\ref{14}), i.e.
\begin{equation}\label{15}
  p'_{i}=p_{i}.
\end{equation}
Then the entropy (\ref{1}), the energy (\ref{5}) and the free energy (\ref{4}) under the energy shift (\ref{14}) are transformed as
\begin{align}\label{16}
  S' = &  -\sum_{i} p'_{i} \ln p'_{i}=  -\sum_{i} p_{i} \ln p_{i} = S, \\ \label{17}
  E' = & \sum_{i}  p'_{i} E'_{i} = \sum_{i}  p_{i} (E_{i} + E_{0}) = E + E_{0}, \\ \label{18}
  F' = & \sum\limits_{i}  p'_{i} (E'_{i} + T \ln p'_{i}) = \sum\limits_{i}  p_{i} (E_{i} + E_{0} + T \ln p_{i}) = F + E_{0}.
\end{align}
Using Eqs.~(\ref{11}), (\ref{14}) and (\ref{16})--(\ref{18}), we obtain
\begin{equation}\label{19}
  p'_{i} = e^{\frac{F'-E'_{i}}{T}} = e^{-S'+\frac{E'-E'_{i}}{T}} = e^{\frac{F-E_{i}}{T}} = e^{-S+\frac{E-E_{i}}{T}} = p_{i}.
\end{equation}
Equation (\ref{19}) is equivalent to Eq.~(\ref{15}). Thus we have proved that the probability distribution $p_{i}$ of the Boltzmann-Gibbs statistics is invariant under an overall shift in energy. Note that the invariance of the distribution (\ref{11}) can be obtained explicitly from the last equality of Eq.~(\ref{11}) and Eq.~(\ref{12}) without introducing the assumption (\ref{15}).

\section{Tsallis-1 statistics}\label{sec2}
The Tsallis-$1$ statistics~\cite{Tsal88} is defined by the generalized entropy with the probabilities normalized to unity~\cite{Tsal88,Tsal98}
\begin{align}\label{20}
    S = & \sum\limits_{i} \frac{p_{i}^{q}-p_{i}}{1-q}, \\ \label{21}
    1= & \sum\limits_{i} p_{i}
\end{align}
and by the standard expectation values
\begin{equation}\label{22}
   \langle A \rangle = \sum\limits_{i} p_{i} A_{i},
\end{equation}
where $q\in\mathbb{R}$ is a real parameter taking values $0<q<\infty$. In the Gibbs limit $q\to 1$, the entropy (\ref{20}) recovers the Boltzmann-Gibbs entropy (\ref{1}) and the Tsallis-1 statistics is reduced to the usual Boltzmann-Gibbs statistics (see Sect.~\ref{sec1}).

Let us consider the canonical ensemble. The thermodynamic potential $F$ of the canonical ensemble can be written as
\begin{align}\label{23}
 F =& E -TS  =  \sum\limits_{i}  p_{i} \left[E_{i} + T \frac{p_{i}^{q-1}-1}{q-1}\right], \\ \label{24}
  E = & \sum_{i}  p_{i} E_{i},
\end{align}
where $E$ is the mean energy and $E_{i}$ is the energy of the $i$-th microscopic state of the system.

Substituting Eq.~(\ref{23}) into Eqs.~(\ref{6})--(\ref{8}) and using Eq.~(\ref{21}), we obtain the equilibrium probabilities of microstates of the system~\cite{Parvan2006a}
\begin{equation}\label{25}
p_{i} = \left[1+\frac{q-1}{q}\frac{\Lambda-E_{i}}{T}\right]^{\frac{1}{q-1}}
\end{equation}
and
\begin{equation}\label{26}
    \sum\limits_{i} \left[1+\frac{q-1}{q}\frac{\Lambda-E_{i}}{T}\right]^{\frac{1}{q-1}}=1,
\end{equation}
where $\Lambda\equiv \lambda-T$ and $\partial E_{i}/\partial p_{i}=0$. In the Gibbs limit $q\to 1$, the probability distribution (\ref{25}) recovers the Boltzmann-Gibbs probability distribution (\ref{9}).  The expectation values of an operator $A$ takes the form~\cite{Parvan2006a}
\begin{equation}\label{27}
   \langle A \rangle = \sum\limits_{i} A_{i} \left[1+\frac{q-1}{q}\frac{\Lambda-E_{i}}{T}\right]^{\frac{1}{q-1}},
\end{equation}
where the norm function $\Lambda$ is the solution of Eq.~(\ref{26}).

Substituting Eq.~(\ref{25}) into Eq.~(\ref{23}) and using Eqs.~(\ref{21}) and (\ref{24}), we obtain~\cite{Parvan2006a}
\begin{equation}\label{28}
  F =  E-TS = \frac{1}{q} [\Lambda + (q-1) E]
\end{equation}
and~\cite{Parvan2006a}
\begin{equation}\label{29}
  \Lambda =  E - q TS.
\end{equation}
Substituting Eq.~(\ref{29}) into Eq.~(\ref{25}), we get (cf. with Eq.~(\ref{11}))
\begin{equation}\label{30}
p_{i} = \left[1 -(q-1)S +\frac{q-1}{q}\frac{E-E_{i}}{T}\right]^{\frac{1}{q-1}}.
\end{equation}
The formula (\ref{30}) was also obtained in ref.~\cite{Sisto} with $\sum_{i} p_{i}^{q}$ instead of $1 -(q-1)S$. From the definition of entropy (\ref{20}) we have $\sum_{i} p_{i}^{q}=1 -(q-1)S$. In the Gibbs limit $q\to 1$, the probability distribution (\ref{30}) of the Tsallis-1 statistics recovers the Boltzmann-Gibbs probability distribution (\ref{11}). Comparing Eqs.~(\ref{30}) and (\ref{11}), we can see that the relation between the energies of microstates, the mean energy and the entropy of system in the probability distribution (\ref{30}) of the Tsallis-1 statistics is consistent with the same relation in the probability distribution (\ref{11}) of the Boltzmann-Gibbs statistics. Thus the probability distribution of the Tsallis-1 statistics has the correct structure from the point of view of the fundamentals of the statistical mechanics. It should be stressed that the probability distribution in terms of the Lagrange multiplier $\beta$ obtained in refs.~\cite{Tsal88,Tsal98,Tsal_book} on the base of the Jaynes principle is precisely reduced to the probability distribution (\ref{25}) (as well as (\ref{30})) after excluding the intermediate Lagrange multiplier $\beta$, which is a method dependent. See the proof in the~\ref{app1}.

Let us demonstrate that the probability distribution of the Tsallis-1 statistics is invariant under the overall energy shift (\ref{14}), $E'_{i}=E_{i} + E_{0}$. Assuming that
\begin{equation}\label{31}
  p'_{i}=p_{i},
\end{equation}
we obtain
\begin{align}\label{32}
   S' = & \sum\limits_{i} \frac{{p'}_{i}^{q}-{p'}_{i}}{1-q} = \sum\limits_{i} \frac{p_{i}^{q}-p_{i}}{1-q} = S, \\ \label{33}
   E'= & \sum\limits_{i} p'_{i} E'_{i} = \sum\limits_{i} p_{i} (E_{i}+E_{0}) = E+E_{0}, \\ \label{34}
   \Lambda'  = & E'-q TS' =  E+E_{0} - q TS = \Lambda +E_{0}
\end{align}
and
\begin{equation}\label{35}
  F' = \sum\limits_{i}  p'_{i} \left[E'_{i} + T \frac{{p'}_{i}^{q-1}-1}{q-1}\right] = \sum\limits_{i}  p_{i} \left[E_{i} + E_{0} + T \frac{p_{i}^{q-1}-1}{q-1}\right] = F+E_{0}.
\end{equation}
Using Eqs.~(\ref{14}), (\ref{25}), (\ref{30}) and (\ref{32})--(\ref{34}), we have (cf. with Eq.~(\ref{19}))
\begin{equation}\label{36}
  p'_{i} = \left[1+\frac{q-1}{q}\frac{\Lambda'-E'_{i}}{T}\right]^{\frac{1}{q-1}} = \left[1+\frac{q-1}{q}\frac{\Lambda-E_{i}}{T}\right]^{\frac{1}{q-1}}= p_{i}
\end{equation}
and
\begin{equation}\label{37}
p'_{i} = \left[1 -(q-1)S' +\frac{q-1}{q}\frac{E'-E'_{i}}{T}\right]^{\frac{1}{q-1}}= \left[1 -(q-1)S +\frac{q-1}{q}\frac{E-E_{i}}{T}\right]^{\frac{1}{q-1}}=p_{i}.
\end{equation}
Equations (\ref{36}) and (\ref{37}) exactly coincide with Eq.~(\ref{31}). Thus we have proved that the probability distribution $p_{i}$ of the Tsallis-1 statistics is indeed invariant under an overall shift in energy.

\section{Tsallis-2 statistics}\label{sec3}
The Tsallis-2 statistics is defined by the generalized entropy (\ref{20}) with probabilities $\{p_{i}\}$ of the microstates of system normalized to unity (\ref{21}) and by the generalized expectation values~\cite{Tsal88,Tsal98,Curado91}
\begin{equation}\label{38}
   \langle A \rangle = \sum\limits_{i} p_{i}^{q} A_{i},
\end{equation}
where $q\in\mathbb{R}$ is a real parameter taking values $0<q<\infty$. This choice of the expectation values is inconsistent because it has the feature $\langle 1 \rangle \neq 1$.

The thermodynamic potential of the canonical ensemble for the Tsallis-$2$ statistics can be written as
\begin{align}\label{39}
 F = & E -TS = \sum\limits_{i}  p_{i}^{q} \left[E_{i} + T \frac{p_{i}^{1-q}-1}{1-q}\right], \\ \label{40}
 E = & \sum_{i}  p_{i}^{q} E_{i},
\end{align}
where $E$ is the mean energy of the system.

Substituting Eq.~(\ref{39}) into Eqs.~(\ref{6})--(\ref{8}) and using Eq.~(\ref{21}), we obtain the equilibrium probabilities of microstates of the system~\cite{Tsal98}
\begin{align}\label{41}
p_{i} =& \frac{1}{Z}  \left[1-(1-q) \frac{E_{i}}{T} \right]^{\frac{1}{1-q}}, \\ \label{42}
    Z =& \sum\limits_{i} \left[1-(1-q) \frac{E_{i}}{T} \right]^{\frac{1}{1-q}},
\end{align}
where
\begin{equation}\label{43}
   Z \equiv \left[1-(1-q)\frac{\Lambda}{T}\right]^{\frac{1}{1-q}}
\end{equation}
and $q\Lambda \equiv \lambda -T$. Then the expectation values (\ref{38}) for the Tsallis-$2$ statistics in the canonical ensemble can be rewritten in the general form as~\cite{Tsal98}
\begin{equation}\label{44}
   \langle A \rangle = \frac{1}{Z^{q}}\sum\limits_{i} A_{i} \left[1-(1-q)\frac{E_{i}}{T}\right]^{\frac{q}{1-q}},
\end{equation}
where the partition function $Z$ is calculated by Eq.~(\ref{42}).

Substituting Eq.~(\ref{41}) into Eq.~(\ref{39}) and using Eqs.~(\ref{43}) and (\ref{20}), and introducing the function $\chi\equiv \sum_{i} p_{i}^{q}=1+(1-q)S$, we obtain
\begin{equation}\label{45}
  F =  E-TS = \Lambda = -T \frac{Z^{1-q}-1}{1-q}.
\end{equation}
Substituting Eq.~(\ref{45}) into Eq.~(\ref{41}), we have (cf. with Eqs.~(\ref{11}) and (\ref{30}))
\begin{equation}\label{46}
p_{i} = \left[\frac{1-(1-q) \frac{E_{i}}{T}}{1 -(1-q)\frac{F}{T}}\right]^{\frac{1}{1-q}} = \left[\frac{1-(1-q) \frac{E_{i}}{T}}{1+(1-q)S -(1-q)\frac{E}{T}}\right]^{\frac{1}{1-q}}.
\end{equation}
Comparing Eqs.~(\ref{46}) and (\ref{11}), we observe that the relation between the energies of microstates, the mean energy and the entropy of system in the probability distribution of the Tsallis-2 statistics is incompatible with the same relation for the probability distribution (\ref{11}) of the Boltzmann-Gibbs statistics. Thus the structure of the probability distribution of the Tsallis-2 statistics does not reproduce the structure of the probability distribution of the Boltzmann-Gibbs statistics. This discrepancy is an additional deficiency of the Tsallis-2 statistics.

Let us verify the invariance of the probability distribution of the Tsallis-2 statistics under the overall energy shift (\ref{14}), $E'_{i}=E_{i} + E_{0}$. Assuming that
\begin{equation}\label{47}
  p'_{i}=p_{i},
\end{equation}
we obtain $S'=S$ (see Eq.~(\ref{32})) and
\begin{align}\label{48}
   E'= & \sum\limits_{i} {p'}_{i}^{q} E'_{i} = \sum\limits_{i} p_{i}^{q} (E_{i}+E_{0}) = E + E_{0} \chi, \\ \label{49}
   F' = & \sum\limits_{i}  {p'}_{i}^{q} \left[E'_{i} + T \frac{{p'}_{i}^{1-q}-1}{1-q}\right] = \sum\limits_{i}  p_{i}^{q} \left[E_{i}+E_{0} + T \frac{p_{i}^{1-q}-1}{1-q}\right] = F+E_{0} \chi.
\end{align}
Using Eqs.~(\ref{14}), (\ref{46}) and (\ref{32}), (\ref{48}), (\ref{49}), we get (cf. with Eq.~(\ref{19}))
\begin{equation}\label{50}
  p'_{i} =\left[\frac{1-(1-q) \frac{E'_{i}}{T}}{1 -(1-q)\frac{F'}{T}}\right]^{\frac{1}{1-q}} = \left[\frac{1-(1-q) \frac{E_{i}+ E_{0}}{T}}{1 -(1-q)\frac{F+E_{0}\chi}{T}}\right]^{\frac{1}{1-q}} \neq p_{i}
\end{equation}
and
\begin{equation}\label{51}
p'_{i} = \left[\frac{1-(1-q) \frac{E'_{i}}{T}}{1+(1-q)S' -(1-q)\frac{E'}{T}}\right]^{\frac{1}{1-q}} = \left[\frac{1-(1-q) \frac{E_{i}+ E_{0}}{T}}{1+(1-q)S -(1-q)\frac{E + E_{0} \chi}{T}}\right]^{\frac{1}{1-q}} \neq p_{i}.
\end{equation}
Equations (\ref{50}) and (\ref{51}) are not equivalent to Eq.~(\ref{47}). Thus the probability distribution $p_{i}$ of the Tsallis-2 statistics is not invariant under an overall shift in energy.

\section{Tsallis-3 statistics}\label{sec4}
The Tsallis-3 statistics is defined by the generalized entropy (\ref{20}) with probabilities $\{p_{i}\}$ of the microstates of system normalized to unity (\ref{21}) and by the normalized generalized expectation values~\cite{Tsal98}
\begin{equation}\label{52}
   \langle A \rangle = \frac{\sum\limits_{i} p_{i}^{q} A_{i}}{\sum\limits_{i} p_{i}^{q}},
\end{equation}
where $q\in\mathbb{R}$ is a real parameter taking values $0<q<\infty$. In contrast with the Tsallis-2 statistics, this choice of the expectation values is consistent because it has the feature $\langle 1 \rangle = 1$.

Let us consider the canonical ensemble. The thermodynamic potential of the canonical ensemble for the Tsallis-$3$ statistics can be written as
\begin{align}\label{53}
 F = & E -TS = \frac{1}{\chi} \sum\limits_{i}  p_{i}^{q} \left[E_{i} + T \chi \frac{p_{i}^{1-q}-1}{1-q}\right], \\ \label{54}
 E = & \frac{1}{\chi} \sum_{i}  p_{i}^{q} E_{i},
\end{align}
where $\chi\equiv \sum_{i} p_{i}^{q}=1+(1-q)S$ and $E$ is the mean energy of the system.

Substituting Eq.~(\ref{53}) into Eqs.~(\ref{6})--(\ref{8}) and using Eq.~(\ref{21}), we obtain the equilibrium probabilities of microstates of the system~\cite{Tsal98}
\begin{align}\label{55}
p_{i} =& \frac{1}{Z}  \left[1+(1-q) \frac{E-E_{i}}{T\chi} \right]^{\frac{1}{1-q}}, \\ \label{56}
    Z =& \sum\limits_{i} \left[1+(1-q) \frac{E-E_{i}}{T\chi} \right]^{\frac{1}{1-q}},
\end{align}
where
\begin{align}\label{57}
   Z \equiv & \left[1-(1-q)\frac{\Lambda}{T}\right]^{\frac{1}{1-q}}, \\ \label{58}
   \chi = & \frac{1}{Z^{q}} \sum\limits_{i} \left[1+(1-q) \frac{E-E_{i}}{T\chi} \right]^{\frac{q}{1-q}}
\end{align}
and $q\Lambda \equiv \lambda -T$. Rewriting Eq.~(\ref{55}) into the form
\begin{equation}\label{59}
  p_{i} Z^{1-q} = p_{i}^{q}  \left[1+(1-q) \frac{E-E_{i}}{T\chi} \right]
\end{equation}
and taking the summation over the index $i$, we obtain~\cite{Tsal98}
\begin{equation}\label{60}
  Z^{1-q} = \chi.
\end{equation}
Substituting Eq.~(\ref{60}) into Eqs.~(\ref{54}) -- (\ref{56}), (\ref{58}) and combining Eqs.~(\ref{56}) and (\ref{58}), we obtain the probability distribution~\cite{Tsal98}
\begin{equation}\label{61}
  p_{i} = \frac{1}{Z}  \left[1+(1-q) \frac{E-E_{i}}{TZ^{1-q}} \right]^{\frac{1}{1-q}}
\end{equation}
and two norm equations
\begin{equation}\label{62}
    \sum\limits_{i} \left[1+(1-q) \frac{E-E_{i}}{TZ^{1-q}} \right]^{\frac{1}{1-q}} =  \sum\limits_{i} \left[1+(1-q) \frac{E-E_{i}}{TZ^{1-q}} \right]^{\frac{q}{1-q}}
\end{equation}
and
\begin{equation}\label{63}
  E =  \frac{1}{Z} \sum\limits_{i} E_{i}  \left[1+(1-q) \frac{E-E_{i}}{TZ^{1-q}} \right]^{\frac{q}{1-q}}
\end{equation}
for two unknown norm functions $E$ and $Z$. Then the expectation values (\ref{52}) for the Tsallis-$3$ statistics in the canonical ensemble can be rewritten in the general form as~\cite{Tsal98}
\begin{equation}\label{64}
   \langle A \rangle = \frac{1}{Z}\sum\limits_{i} A_{i} \left[1+(1-q) \frac{E-E_{i}}{TZ^{1-q}} \right]^{\frac{q}{1-q}},
\end{equation}
where the norm functions $Z$ and $E$ are calculated from the system of two equations (\ref{62}) and (\ref{63}). Note that Eq.~(\ref{60}) (as well as Eq.~(\ref{62})) represents the equation of self-consistency of the Tsallis-3 statistics.

Substituting Eq.~(\ref{61}) into Eqs.~(\ref{20}) and (\ref{53}), and using Eq.~(\ref{60}), we obtain~\cite{Tsal98}
\begin{align}\label{65}
   S = & \frac{Z^{1-q}-1}{1-q}, \\ \label{66}
   F = & E-TS = E - T \frac{Z^{1-q}-1}{1-q}.
\end{align}
Substituting Eq.~(\ref{65}) into Eq.~(\ref{61}), we have (cf. with Eqs.~(\ref{11}) and (\ref{30}))
\begin{equation}\label{66a}
  p_{i} = \left[\frac{1}{1+(1-q)S}+(1-q) \frac{E-E_{i}}{T(1+(1-q)S)^{2}} \right]^{\frac{1}{1-q}}.
\end{equation}
Comparing Eqs.~(\ref{66a}) and (\ref{11}), we observe that the relation between the energies of microstates, the mean energy and the entropy of system in the probability distribution (\ref{66a}) of the Tsallis-3 statistics is not compatible with the same relation for the probability distribution (\ref{11}) of the Boltzmann-Gibbs statistics. Thus the probability distribution of the Tsallis-3 statistics does not reproduce the structure of the probability distribution of the Boltzmann-Gibbs statistics, which is correct from the point of view of the fundamentals of the statistical mechanics. This discrepancy is a significant deficiency of the Tsallis-3 statistics.

Let us prove that the probability distribution of the Tsallis-3 statistics is invariant under the overall energy shift (\ref{14}), $E'_{i}=E_{i} + E_{0}$. Assuming that
\begin{equation}\label{67}
  p'_{i}=p_{i},
\end{equation}
we obtain $S'=S$ (see Eq.~(\ref{32})) and
\begin{align}\label{68}
   \chi' = & \sum\limits_{i} {p'}_{i}^{q} = \sum\limits_{i} p_{i}^{q} = \chi, \\ \label{69}
   Z'= & \left(\chi' \right)^{\frac{1}{1-q}} = \left(\chi \right)^{\frac{1}{1-q}} = Z, \\ \label{70}
   E'= & \frac{1}{\chi'} \sum\limits_{i} {p'}_{i}^{q} E'_{i} = \frac{1}{\chi} \sum\limits_{i} p_{i}^{q} (E_{i}+E_{0}) = E+E_{0}
\end{align}
and
\begin{equation}\label{71}
  F' = E'-TS'= E+E_{0} -TS  = F + E_{0}.
\end{equation}
Using Eqs.~(\ref{14}), (\ref{61}), (\ref{69}) and (\ref{70}), we have (cf. with Eq.~(\ref{19}))
\begin{equation}\label{72}
   p'_{i} = \frac{1}{Z'}  \left[1+(1-q) \frac{E'-E'_{i}}{T{Z'}^{1-q}} \right]^{\frac{1}{1-q}} = \frac{1}{Z}  \left[1+(1-q) \frac{E-E_{i}}{TZ^{1-q}} \right]^{\frac{1}{1-q}} = p_{i}.
\end{equation}
Equation (\ref{72}) coincides with Eq.~(\ref{67}). Thus we have proved that the probability distribution $p_{i}$ of the Tsallis-3 statistics is invariant under an overall shift in energy.

\section{$q$-dual statistics}\label{sec5}
The $q$-dual statistics is defined by the generalized entropy obtained from the entropy (\ref{20}) by the multiplicative transformation $q\to 1/q$ with the probabilities normalized to unity (see the reference~\cite{Parvan2020b})
\begin{align}\label{73}
    S = & \frac{q}{q-1}\sum\limits_{i} \left(p_{i}^{1/q}-p_{i}\right), \\ \label{74}
    1= & \sum\limits_{i} p_{i}
\end{align}
and by the standard expectation values
\begin{equation}\label{75}
   \langle A \rangle = \sum\limits_{i} p_{i} A_{i},
\end{equation}
where $q\in\mathbb{R}$ is a real parameter taking values $0<q<\infty$. In the Gibbs limit $q\to 1$, the entropy (\ref{73}) recovers the Boltzmann-Gibbs entropy (\ref{1}) and the $q$-dual statistics is reduced to the usual Boltzmann-Gibbs statistics (see Sect.~\ref{sec1}). Note that the $q$-dual statistics under the transformation $q\to 1/q$ is reduced to the Tsallis-1 statistics (see Sect.~\ref{sec2}). It should be stressed that the $q$-dual statistics was especially introduced in physics in ref.~\cite{Parvan2020b} in order to justify the phenomenological Tsallis transverse momentum distribution~\cite{Cleymans2012,Cleymans12a}, which is largely used in the high energy physics.

Let us consider the canonical ensemble. The thermodynamic potential $F$ of the canonical ensemble can be written as
\begin{align}\label{76}
 F =& E -TS  =  \sum\limits_{i}  p_{i} \left[E_{i} + T q \frac{p_{i}^{\frac{1-q}{q}}-1}{1-q}\right], \\ \label{77}
  E = & \sum_{i}  p_{i} E_{i},
\end{align}
where $E$ is the mean energy and $E_{i}$ is the energy of the $i$-th microscopic state of the system.

Substituting Eq.~(\ref{76}) into Eqs.~(\ref{6})--(\ref{8}) and using Eq.~(\ref{74}), we obtain the equilibrium probabilities of microstates of the system
\begin{equation}\label{78}
p_{i} = \left[1+(1-q)\frac{\Lambda-E_{i}}{T}\right]^{\frac{q}{1-q}}
\end{equation}
and
\begin{equation}\label{79}
    \sum\limits_{i}  \left[1+(1-q)\frac{\Lambda-E_{i}}{T}\right]^{\frac{q}{1-q}} = 1,
\end{equation}
where $\Lambda\equiv \lambda-T$ and $\partial E_{i}/\partial p_{i}=0$. In the Gibbs limit $q\to 1$, the probability distribution (\ref{78}) recovers the Boltzmann-Gibbs probability distribution (\ref{9}).  The expectation values of an operator $A$ takes the form
\begin{equation}\label{80}
   \langle A \rangle = \sum\limits_{i} A_{i}  \left[1+(1-q)\frac{\Lambda-E_{i}}{T}\right]^{\frac{q}{1-q}},
\end{equation}
where the norm function $\Lambda$ is the solution of Eq.~(\ref{79}).

Substituting Eq.~(\ref{78}) into Eq.~(\ref{76}) and using Eqs.~(\ref{74}) and (\ref{77}), we obtain
\begin{equation}\label{81}
  F =  E-TS = q \left[\Lambda + \frac{1-q}{q} E\right]
\end{equation}
and
\begin{equation}\label{82}
  \Lambda =  E - \frac{1}{q} TS.
\end{equation}
Substituting Eq.~(\ref{82}) into Eq.~(\ref{78}), we have (cf. with Eqs.~(\ref{11}) and (\ref{30}))
\begin{equation}\label{83}
p_{i} = \left[1 -\frac{1-q}{q} S +(1-q)\frac{E-E_{i}}{T}\right]^{\frac{q}{1-q}}.
\end{equation}
In the Gibbs limit $q\to 1$, the probability distribution (\ref{83}) of the $q$-dual statistics recovers the Boltzmann-Gibbs probability distribution (\ref{11}). Comparing Eqs.~(\ref{83}) and (\ref{11}), we observe that the relation between the energies of microstates, the mean energy and the entropy of system in the probability distribution (\ref{83}) of the $q$-dual statistics is compatible with the same relation for the probability distribution (\ref{11}) of the Boltzmann-Gibbs statistics. Thus the probability distribution of the $q$-dual statistics has the correct structure.

Let us prove that the probability distribution of the $q$-dual statistics is invariant under the overall energy shift (\ref{14}), $E'_{i}=E_{i} + E_{0}$. Assuming that
\begin{equation}\label{84}
  p'_{i}=p_{i},
\end{equation}
we obtain
\begin{align}\label{85}
   S' = & \frac{q}{q-1}\sum\limits_{i} \left({p'}_{i}^{1/q}-p'_{i}\right) = \frac{q}{q-1}\sum\limits_{i} \left(p_{i}^{1/q}-p_{i}\right) = S, \\ \label{86}
   E'= & \sum\limits_{i} p'_{i} E'_{i} = \sum\limits_{i} p_{i} (E_{i}+E_{0}) = E+E_{0}, \\ \label{87}
   \Lambda'  = & E'- \frac{1}{q} TS' =  E+E_{0} - \frac{1}{q} TS = \Lambda +E_{0}
\end{align}
and
\begin{equation}\label{88}
  F' = \sum\limits_{i}  p'_{i} \left[E'_{i} + T q \frac{{p'}_{i}^{\frac{1-q}{q}}-1}{1-q}\right] = \sum\limits_{i}  p_{i} \left[E_{i} +E_{0} + T q \frac{p_{i}^{\frac{1-q}{q}}-1}{1-q}\right] = F+E_{0}.
\end{equation}
Using Eqs.~(\ref{14}), (\ref{78}), (\ref{83}) and (\ref{85})--(\ref{87}), we obtain (cf. with Eq.~(\ref{19}))
\begin{equation}\label{89}
  p'_{i} = \left[1+(1-q)\frac{\Lambda'-E'_{i}}{T}\right]^{\frac{q}{1-q}} = \left[1+(1-q)\frac{\Lambda-E_{i}}{T}\right]^{\frac{q}{1-q}}= p_{i}
\end{equation}
and
\begin{equation}\label{90}
p'_{i} = \left[1 -\frac{1-q}{q} S' +(1-q)\frac{E'-E'_{i}}{T}\right]^{\frac{q}{1-q}}= \left[1 -\frac{1-q}{q} S +(1-q)\frac{E-E_{i}}{T}\right]^{\frac{q}{1-q}}=p_{i}.
\end{equation}
Equations (\ref{89}) and (\ref{90}) exactly coincide with Eq.~(\ref{84}). Thus we have proved that the probability distribution $p_{i}$ of the $q$-dual statistics is indeed invariant under an overall shift in energy.

\section{Conclusions}\label{sec6}
We have considered the general formalisms of the Boltzmann-Gibbs statistics, the $q$-dual statistics, the Tsallis-1 statistics, the Tsallis-2 statistics, and the Tsallis-3 statistics in the canonical ensemble. The equilibrium probability distributions have been analytically derived from a constrained local extremum of the thermodynamic potential. The canonical ensemble for the $q$-dual statistics has been studied for the first time. We have found that the probability distributions of all these statistics may be written in different representations. For the probability distributions of the Tsallis-1 statistics, the Tsallis-3 statistics, the $q$-dual statistics, and Boltzmann-Gibbs statistics, we have obtained a representation in which the probability distribution is a function of the difference between the mean energy and the energy of the microstate. However, for the Tsallis-2 statistics, such a representation does not exist. We have found that the probability distributions of the Tsallis-1 and $q$-dual statistics have the correct structure from the point of view of the fundamentals of the statistical mechanics since the relation between the energies of microstates, the mean energy, and the entropy of the system in these probability distributions is consistent with the same relation in the probability distribution of the Boltzmann-Gibbs statistics. However, the probability distributions of the Tsallis-2 and Tsallis-3 statistics do not reproduce this structure of the probability distribution of the Boltzmann-Gibbs statistics, which is correct from the point of view of the fundamentals of the statistical mechanics. Thus this discrepancy is a significant deficiency of the Tsallis-2 and Tsallis-3 statistics.

We have found that the probability distribution of the Tsallis-1 statistics in terms of the Lagrange multiplier $\beta$ obtained in the Tsallis' original paper on the base of the Jaynes principle is precisely reduced to the probability distribution obtained in the present paper and the previous paper of the same author. It has been obtained after excluding the intermediate Lagrange multiplier $\beta$ and representing the final results in terms of the thermodynamic temperature of the system. Thus we have proved the equivalence of the Jaynes principle with the method of extremization of the thermodynamic potential. This fact demonstrates that the final results for the probability distribution of the Tsallis-1 statistics are independent of the method of calculation.

We have introduced a new general and explicit method to demonstrate the invariance of the probability distributions under the uniform energy spectrum translation. This method does not depend on the factorization property of the exponential function. Using this method, we have studied the invariance of the probability distributions of the Boltzmann-Gibbs statistics, the $q$-dual statistics, and three versions of the Tsallis statistics under an overall shift in energy at a fixed temperature. We have demonstrated that the probability distribution of the Tsallis-1 statistics is invariant under the uniform energy spectrum translation at a fixed temperature. We also prove the invariance of the probability distributions of the Tsallis-3 statistics, the $q$-dual statistics, and Boltzmann-Gibbs statistics. However, we have found that the probability distribution of the Tsallis-2 statistics, the expectation values of which are not consistent with the normalization condition of probabilities, is not invariant under the overall shift in energy. Thus this fact presents a severe deficiency for the Tsallis-2 statistics as it violates the fundamentals of physics.

\vskip0.2in
\noindent
{\bf Acknowledgments:}  This work was supported in part by the joint research project of JINR and IFIN-HH.

\vskip0.2in
\appendix

\section{Probability distribution of Tsallis-1 statistics from the Jaynes principle}\label{app1}
Let us show that the probability distribution obtained in the original paper of C.~Tsallis \cite{Tsal88} in terms of the Lagrange multiplier $\beta$ associated with the energy represents an intermediate result and in the terms of temperature $T$ it is reduced to the probability distribution obtained in ref.~\cite{Parvan2006a} and present paper.

In ref.~\cite{Tsal88}, the Tsallis-1 statistics is defined by the generalized entropy obtained from Eq.~(\ref{20}) with the probabilities normalized to unity
\begin{align}\label{a1}
    S = &  \frac{1 - \sum\limits_{i} p_{i}^{q}}{q-1}, \\ \label{a2}
    1= & \sum\limits_{i} p_{i}
\end{align}
and by the standard expectation values
\begin{equation}\label{a3}
   \langle A \rangle = \sum\limits_{i} p_{i} A_{i},
\end{equation}
where $q\in\mathbb{R}$ is a real parameter taking values $0<q<\infty$.

In ref.~\cite{Tsal88}, the Lagrange function of canonical ensemble is defined as
\begin{equation}\label{a4}
 \Phi = S + \alpha \sum\limits_{i} p_{i} - \alpha\beta (q-1) \sum\limits_{i} p_{i} E_{i},
\end{equation}
where $\alpha$ and $\beta$ are the Lagrange multipliers. Imposing $\partial \Phi/\partial p_{i}=0$, one obtains the equilibrium probabilities of microstates of the system
\begin{align}\label{a5}
    p_{i} = &  \frac{1}{Z} \left[1-\beta (q-1) E_{i}\right]^{\frac{1}{q-1}}, \\ \label{a6}
    Z = & \sum\limits_{i} \left[1-\beta (q-1) E_{i}\right]^{\frac{1}{q-1}},
\end{align}
where $Z=(\alpha (q-1)/q)^{-1/(q-1)}$.

Substituting Eq.~(\ref{a5}) into Eq.~(\ref{a1}), we have
\begin{align}\label{a7}
  S= & \frac{1}{q-1} + \frac{Z^{1-q}}{1-q} \left[1-\beta (q-1) E \right], \\ \label{a8}
  E = & \sum\limits_{i} p_{i} E_{i},
\end{align}
where $E$ is the mean energy. Differentiating $S$, $p_{i}$ and $E$ with respect to $\beta$, we obtain
\begin{align}\label{a9}
  \frac{\partial S}{\partial \beta} = & Z^{1-q} \left\{\frac{\partial \ln Z}{\partial \beta} [1-\beta (q-1) E] + E +\beta \frac{\partial E}{\partial \beta} \right\}, \\ \label{a10}
  \frac{\partial p_{i}}{\partial \beta}  = & - \frac{\partial \ln Z}{\partial \beta} p_{i} -  \frac{1}{Z} E_{i} \left[1-\beta (q-1) E_{i}\right]^{\frac{1}{q-1}-1}, \\ \label{a11}
  \frac{\partial E}{\partial \beta} = & \sum\limits_{i} \frac{\partial p_{i}}{\partial \beta} E_{i},
\end{align}
where $\partial E_{i}/\partial \beta=0$. Multiplying Eq.~(\ref{a10}) by the factor $[1-\beta (q-1) E_{i}]$ and taking the summation over $i$, and using Eqs.~(\ref{a2}), (\ref{a8}), (\ref{a11}) and the equation $\sum\limits_{i} \partial p_{i}/\partial \beta=0$, we obtain
\begin{equation}\label{a12}
  -\frac{\partial \ln Z}{\partial \beta} =  \frac{E-\beta (q-1)\frac{\partial E}{\partial \beta}}{1-\beta (q-1) E}
\end{equation}
and
\begin{equation}\label{a13}
  \frac{\partial S}{\partial \beta}  = q \beta Z^{1-q} \frac{\partial E}{\partial \beta}.
\end{equation}
The temperature in the statistical mechanics is defined from the fundamental thermodynamic potential $E$ as
\begin{equation}\label{a14}
  T\equiv  \frac{\partial E}{\partial S} = \frac{\partial E/\partial \beta}{\partial S/\partial \beta}.
\end{equation}
Substituting Eq.~(\ref{a13}) into Eq.~(\ref{a14}), we obtain
\begin{equation}\label{a15}
  \beta =\frac{Z^{q-1}}{qT}.
\end{equation}
Thus the Lagrange multiplier $\beta$ is a function of the thermodynamic temperature $T$.

Substituting Eq.~(\ref{a15}) into Eqs.~(\ref{a5})--(\ref{a7}), we have
\begin{align}\label{a16}
    p_{i} = &  \frac{1}{Z} \left[1- \frac{q-1}{q} \frac{E_{i}}{T Z^{1-q}} \right]^{\frac{1}{q-1}}, \\ \label{a17}
    Z = & \sum\limits_{i} \left[1-\frac{q-1}{q} \frac{E_{i}}{T Z^{1-q}} \right]^{\frac{1}{q-1}}
\end{align}
and
\begin{align}\label{a18}
  S= & \frac{1-Z^{1-q}}{q-1} + \frac{1}{q} \frac{E}{T}, \\ \label{a19}
  F = & E-TS = -T \frac{1-Z^{1-q}}{q-1} + \frac{q-1}{q} E.
\end{align}
Using Eqs.~(\ref{a16}), (\ref{a18}), we obtain exactly Eq.~(\ref{30}) of the Tsallis-1 statistics:
\begin{equation}\label{a20}
p_{i} = \left[1 -(q-1)S +\frac{q-1}{q}\frac{E-E_{i}}{T}\right]^{\frac{1}{q-1}}.
\end{equation}

Let us introduce a new function $\Lambda$ instead of $Z$ in the form
\begin{equation}\label{a21}
    Z^{1-q}\equiv 1+\frac{q-1}{q}\frac{\Lambda}{T},
\end{equation}
Substituting Eq.~(\ref{a21}) into Eqs.~(\ref{a16}), (\ref{a17}), we obtain exactly Eqs.~(\ref{25}) and (\ref{26}) of the Tsallis-1 statistics:
\begin{equation}\label{a22}
p_{i} = \left[1+\frac{q-1}{q}\frac{\Lambda-E_{i}}{T}\right]^{\frac{1}{q-1}}
\end{equation}
and
\begin{equation}\label{a23}
    \sum\limits_{i} \left[1+\frac{q-1}{q}\frac{\Lambda-E_{i}}{T}\right]^{\frac{1}{q-1}}=1.
\end{equation}
Substituting Eq.~(\ref{a21}) into Eqs.~(\ref{a18}), (\ref{a19}), we obtain exactly Eqs.~(\ref{28}) and (\ref{29}) of the Tsallis-1 statistics:
\begin{equation}\label{a24}
  F =  E-TS = \frac{1}{q} [\Lambda + (q-1) E]
\end{equation}
and
\begin{equation}\label{a25}
  \Lambda =  E - q TS.
\end{equation}
Thus the formalism of the Tsallis-1 statistics given in refs.~\cite{Tsal88,Tsal98} is reduced to the formalism obtained in ref.~\cite{Parvan2006a} and present paper after excluding the intermediate Lagrange multiplier $\beta$, which is a method dependent. Note that this proof in a concise form can be found also in ref.~\cite{Parvan2006a}.





\end{document}